\documentclass[letterpaper,twocolumn,prl,aps,superscriptaddress,amsmath,amssymb,floatfix]{revtex4-1}
\usepackage{mathptmx}
\usepackage[latin9]{inputenc}
\usepackage{color}
\usepackage{amsmath}
\usepackage{amssymb}
\usepackage{graphicx}
\usepackage{esint}
\usepackage[unicode=true,
bookmarks=true,bookmarksnumbered=false,bookmarksopen=false,
breaklinks=false,pdfborder={0 0 1},backref=false,colorlinks=true]
{hyperref}
\hypersetup{
	linkcolor=magenta,urlcolor=blue,citecolor=blue,pdfstartview={FitH},hyperfootnotes=false}

\makeatletter

\usepackage{textcomp}
\usepackage{epstopdf}

\usepackage{amsfonts}

\pdfpageheight\paperheight
\pdfpagewidth\paperwidth

\@ifundefined{textcolor}{}{%
	\definecolor{BLACK}{gray}{0}
	\definecolor{WHITE}{gray}{1}
	\definecolor{RED}{rgb}{1,0,0}
	\definecolor{GREEN}{rgb}{0,1,0}
	\definecolor{BLUE}{rgb}{0,0,1}
	\definecolor{CYAN}{cmyk}{1,0,0,0}
	\definecolor{MAGENTA}{cmyk}{0,1,0,0}
	\definecolor{YELLOW}{cmyk}{0,0,1,0}
}

\usepackage{xcolor}\usepackage{soul}
\newcommand{\ket}[1]{\ensuremath{\left|#1\right\rangle}}

\definecolor{blue}{rgb}{0,0,1}
\definecolor{red}{rgb}{1,0,0}
\definecolor{green}{rgb}{0,1,0}

\makeatother

\begin{document}
	\title{Error-transparent operations on a logical qubit protected by quantum
		error correction}
	\author{Y.~Ma}
	\affiliation{Center for Quantum Information, Institute for Interdisciplinary Information
		Sciences, Tsinghua University, Beijing 100084, China}
	\author{Y.~Xu}
	\affiliation{Center for Quantum Information, Institute for Interdisciplinary Information
		Sciences, Tsinghua University, Beijing 100084, China}
	\author{X.~Mu}
	\affiliation{Center for Quantum Information, Institute for Interdisciplinary Information
		Sciences, Tsinghua University, Beijing 100084, China}
	\author{W.~Cai}
	\affiliation{Center for Quantum Information, Institute for Interdisciplinary Information
		Sciences, Tsinghua University, Beijing 100084, China}
	\author{L.~Hu}
	\affiliation{Center for Quantum Information, Institute for Interdisciplinary Information
		Sciences, Tsinghua University, Beijing 100084, China}
	\author{W.~Wang}
	\affiliation{Center for Quantum Information, Institute for Interdisciplinary Information
		Sciences, Tsinghua University, Beijing 100084, China}
	\author{X.~Pan}
	\affiliation{Center for Quantum Information, Institute for Interdisciplinary Information
		Sciences, Tsinghua University, Beijing 100084, China}
	\author{H.~Wang}
	\affiliation{Center for Quantum Information, Institute for Interdisciplinary Information
		Sciences, Tsinghua University, Beijing 100084, China}
	\author{Y.~P.~Song}
	\affiliation{Center for Quantum Information, Institute for Interdisciplinary Information
		Sciences, Tsinghua University, Beijing 100084, China}
	\author{C.-L.~Zou}
	\email{clzou321@ustc.edu.cn}
	
	\affiliation{Key Laboratory of Quantum Information, CAS, University of Science
		and Technology of China, Hefei, Anhui 230026, P. R. China}
	\author{L.~Sun}
	\email{luyansun@tsinghua.edu.cn}
	
	\affiliation{Center for Quantum Information, Institute for Interdisciplinary Information
		Sciences, Tsinghua University, Beijing 100084, China}
	%\date{\today}
	
	\begin{abstract}
		Universal quantum computation~\cite{Nielsen} is striking
		for its unprecedented capability in processing information, but its
		scalability is challenging in practice because of the inevitable environment
		noise. Although quantum error correction (QEC) techniques~\cite{Chiaverini2004,Schindler2011,Reed2012,Yao2012,Waldherr2014,Kelly2015} have been developed to protect stored quantum information from leading
		orders of errors, the noise-resilient processing of the QEC-protected
		quantum information is highly demanded but remains elusive~\cite{Campbell2017}.
		Here, we demonstrate phase gate operations on a logical qubit encoded
		in a bosonic oscillator in an error-transparent (ET) manner. Inspired
		by Refs.~\cite{vy2013error,kapit2018error}, the ET gates are extended
		to the bosonic code and are able to tolerate errors during the gate operations,
		regardless of the random occurrence time of the error. With precisely
		designed gate Hamiltonians through photon-number-resolved AC-Stark
		shifts, the ET condition is fulfilled experimentally. We verify that
		the ET gates outperform the non-ET gates with a substantial improvement of the gate fidelity after an occurrence of the single-photon-loss error. Our ET
		gates in the superconducting quantum circuits are readily for extending
		to multiple encoded qubits and a universal gate set is within reach,
		paving the way towards fault-tolerant quantum computation.
	\end{abstract}
	\maketitle
	
	%about twice longer lifetime of the encoded quantum information under repetitive autonomous QEC protection compared to the case without QEC
	
	%The traditional QEC approach using multiple physical qubits to encode one logical qubit remains very challenging because of the extremely low gate error threshold and the large resources required for concatenated encoding. On the other hand, surface code~\cite{Fowler2012} tolerates a much higher error rate, but at the cost of huge resource overhead.
	
	The uncontrollable noise in a quantum system is the most significant
	obstacle in realizing universal quantum computation~\cite{Nielsen},
	since the induced errors are unpredictable and deleterious to the
	encoded quantum information. Quantum error correction (QEC) is proposed
	to tackle this problem~\cite{Shor1995} by expanding the dimension
	of the Hilbert space for quantum information and thus introducing
	the redundancy to tolerate the leading errors. In conventional QEC,
	quantum information is encoded on logical qubits, constructed by multiple
	physical qubits, within a subspace spanned by the QEC codewords called
	the code space. Although each physical qubit is susceptible to noise,
	errors can be detected without corrupting the stored quantum information
	while mapping the quantum state in the code space to the orthogonal
	error spaces. Over the past years, great progress has been achieved
	in QEC theories, and proof-of-principle demonstrations of error detection
	and correction are reported in various experimental platforms~\cite{Chiaverini2004,Schindler2011,Reed2012,Yao2012,Waldherr2014,Kelly2015}.
	Especially, the break-even point of QEC has been demonstrated with
	a logical qubit encoded in a bosonic oscillator~\cite{ofek2016extending}.
	
	%[width=0.45\textwidth]
	\begin{figure}
		\centering{}\includegraphics{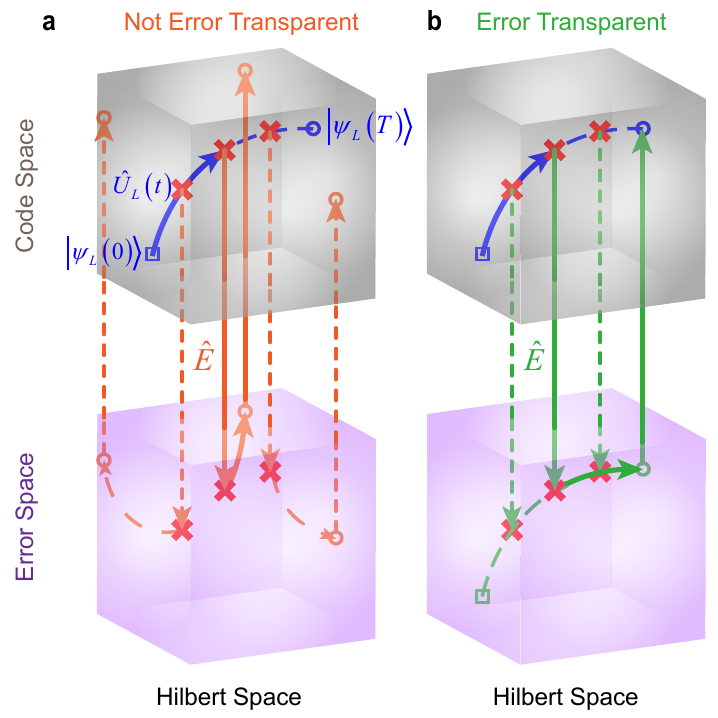} \caption{\textbf{Concept of error-transparent (ET) gate.} During a quantum
			gate on a logical state with finite operation time, error $E$
			might occur randomly, mapping the state in the code space (upper gray
			cube) to the orthogonal error space (lower purple cube). For the non-ET
			gate (\textbf{a}), the track of state evolution in the error space is different
			for $E$ occurring at different time. When the ET condition
			is satisfied (\textbf{b}), the tracks in both the code and error spaces are deterministic and identical whenever the error occurs. Therefore, $E$ can
			be detected and corrected with QEC after the gate operation, making
			the ET gate fault tolerant.}
		\label{fig:ETconcept}
	\end{figure}
	
	However, QEC can merely maintain the stored quantum states from
	noise. Errors occurring during the execution of quantum operations
	might accumulate and spread over the quantum circuits, so that the
	processing of information is not reliable. Fault-tolerant universal quantum computation architectures~\cite{Campbell2017}, such as the transversal gates
	on logical qubits and magic-state distillation, were developed for
	performing noise-resilient quantum gates on encoded qubits, but the
	implementations are extremely challenging. Instead of realizing a
	complete fault-tolerant architecture, practical schemes that demonstrate
	the key ideas in a near-term few-qubit system were proposed~\cite{Gottesman2016,Chao2018}. Only very recently, fault-tolerant state preparation~\cite{Takita2017} and error detection~\cite{Linke2017,Rosenblum2018} were experimentally demonstrated. An alternative approach of fault-tolerant operations based on the concept of error-transparent (ET) gates~\cite{vy2013error,kapit2018error} was proposed theoretically and promises fault-tolerant non-Clifford logical gates. Nevertheless, its implementation in the multi-qubit QEC codes requires many-body interactions and is hard to realize experimentally.
	
	Here, we extend the concept of ET gates to bosonic QEC codes and experimentally
	demonstrate ET arbitrary phase gates that tolerate the single-photon-loss
	error. The ET gates are successfully validated by the remarkable improvement
	of the coherence of the logical states after the occurrence of an error
	during the evolution of the gates. By applying repetitive autonomous QEC
	(AQEC), the ET gates on the QEC-protected logical qubits show higher
	fidelities than the cases with non-ET gates or without AQEC. Our results
	promise a universal ET gate set for quantum computation, thus reveals
	the potential of the bosonic quantum computation architecture~\cite{Gao2019,Fluhmann2019}, and
	presents a first step towards the fault-tolerant quantum computation.
	
	%[width=0.8\textwidth]
	\begin{figure*}
		\centering{}\includegraphics{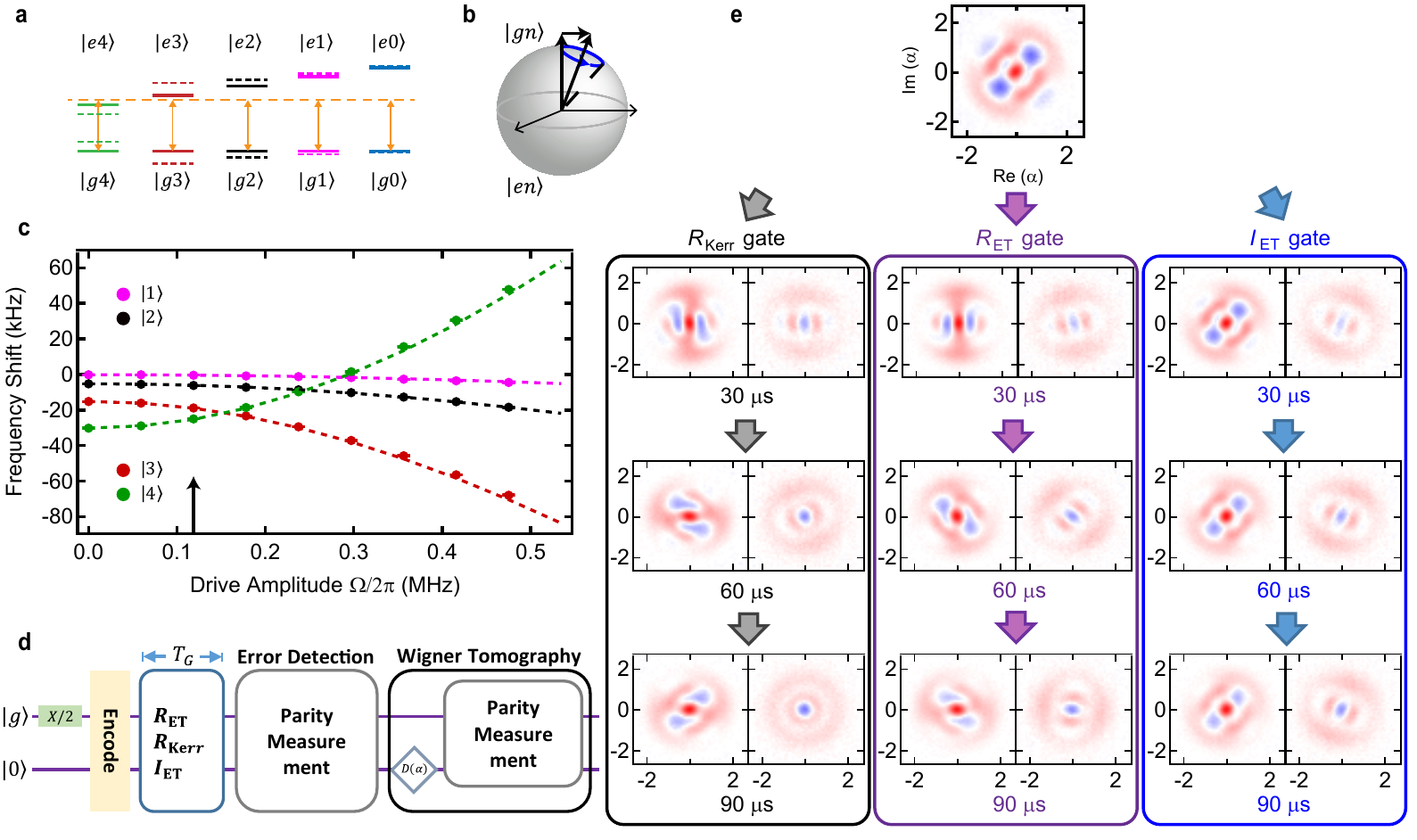} \caption{\textbf{Photon-number-resolved AC-Stark shift (PASS) and ET phase
				gates.} \textbf{a,} Illustration of the energy diagram in the strongly
			dispersively coupled bosonic mode-ancilla system. A detuned microwave
			drive on the ancilla would induce PASS for each transition frequency.
			\textbf{b,} Geometric phase interpretation of the PASS. Fast rotating
			of the state in the Bloch space $\{\ket{gn},\ket{en}\}$ induces a
			phase accumulation proportional to time, equivalent to an energy shift
			of Fock state $\ket{n}$. \textbf{c,} Measured frequencies
			of Fock states with respect to $|0\rangle$ as a function of the
			amplitude of a continuous drive. The drive frequency is in the middle
			of the ancilla transition frequencies corresponding to Fock states
			$|3\rangle$ and $|4\rangle$, as indicated by the horizontal dashed
			orange line in \textbf{a}. These results fit excellently with the
			theoretical predictions (dashed lines). The vertical arrow indicates
			the amplitude to realize the ET phase gate $R_{\mathrm{ET}}$. \textbf{d,}
			Experimental sequence to characterize the ET gates. After $R_{\mathrm{Kerr}}$,
			$R_{\mathrm{ET}}$, or $I_{\mathrm{ET}}$, an error detection measurement
			is performed, followed by a Wigner tomography. \textbf{e,} Evolution
			of the logical state encoded with the lowest-order binomial code $(|0_{L}\rangle-i|1_{L}\rangle)/\sqrt{2}$
			under the three different gates for different intervals of $30~\mu$s,
			$60~\mu$s, and $90~\mu$s. Each left column is the Wigner function
			for a detection of no error, while each right one for a detection
			of an error. For non-ET $R_{\mathrm{Kerr}}$, the state after an error
			eventually loses the phase information and becomes a mixed state;
			while for the ET gates, the coherence of the state in the error space
			is preserved. All the Wigner functions are experimentally measured ones.}
		\label{fig:wignerFig}
	\end{figure*}
	
	The basic idea of the ET operation on a logical qubit is illustrated
	in Fig.~\ref{fig:ETconcept}. Applying a Hamiltonian $H(t)$ to the
	logical qubit, any encoded quantum state is expected to evolve as
	$|\psi_{\mathrm{L}}\left(t_{1}\right)\rangle\rightarrow U\left(t_{2},t_{1}\right)|\psi_{\mathrm{L}}\left(t_{1}\right)\rangle$ with a target unitary operation $U\left(t_{2},t_{1}\right)=\displaystyle{\mathcal{T}}e^{-i\int_{t_{1}}^{t_{2}}d\tau H\left(\tau\right)}$ ($\displaystyle{\mathcal{T}}$ is the time-ordering operator).
	However, because errors could occur during the operation process,
	the logical state will consequently jump from the code space to the
	error space. Due to the stochastic nature of noise, the practical
	evolution of the state may follow different tracks, as shown in Fig.~\ref{fig:ETconcept}\textbf{a}. For example, if an error $E_{j}$ occurs at time $t$, the track leads to the final state $|\widetilde{\psi}(t)\rangle=U(T,t)E_{j}U(t,0)|\psi_{\mathrm{L}}(0)\rangle$
	in the error space. An ET operation requires a deterministic track of
	the logical state evolution irrespective to $t$, as shown in Fig.~\ref{fig:ETconcept}\textbf{b}, and the target operation could always be achieved by mapping the state back to the code space after the operation.
	
	Therefore, we derive the condition for the ET operation as $U(T,t)E_{j}U(t,0)|\psi_{\mathrm{L}}(0)\rangle=e^{i\phi(t)}E_{j}U(T,0)|\psi_{\mathrm{L}}(0)\rangle,\forall j,t,\left|\psi_{\mathrm{L}}(0)\right\rangle$ by considering the
	fact that a global phase $\phi(t)$ makes no influence on the logical
	state. Note that the ET condition discussed in Refs.~\cite{vy2013error,kapit2018error}
	is a more restricted case of our condition with $\phi\left(t\right)=0$.
	Accordingly, the Hamiltonian should be carefully engineered to make
	the evolution in the error spaces exactly the same as that in the
	code space, i.e.
	\begin{equation}
	\mathcal{P}_{j}^{\dagger}H(t)\mathcal{P}_{j}=\mathcal{P}_{\mathrm{C}}^{\dagger}H(t)\mathcal{P}_{\mathrm{C}}+c(t)\mathcal{P}_{\mathrm{C}},\forall j,t,\label{eq:ETcondition2}
	\end{equation}
	where $\mathcal{P}_{\mathrm{C}}$ is the projector onto the code space,
	$\mathcal{P}_{j}\propto E_{j}\mathcal{P}_{\mathrm{C}}$ is the projector
	from the code space to the error space corresponding to the error $E_{j}$,
	and $c(t)$ is a complex number.
	
	%[width=0.8\textwidth]
	\begin{figure*}
		\centering{}\includegraphics{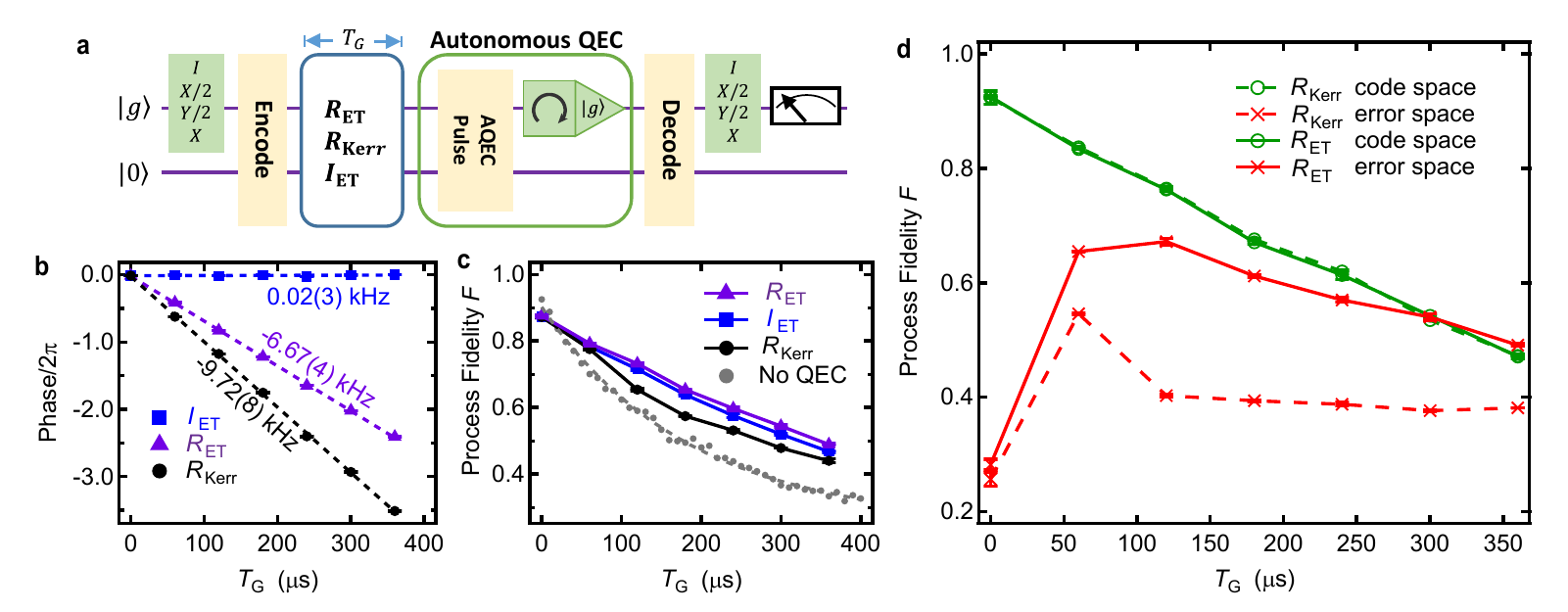} \caption{\textbf{ET gates on a logical qubit with autonomous quantum error
				correction (AQEC).} \textbf{a,} Experimental procedure for the ET
			gate performance characterization. The gate operation on the logical
			qubit is realized by the corresponding Hamiltonian with varying gate
			time $T_{G}$, and a QEC process is implemented before the decoding.
			The QEC process consists of an AQEC pulse followed by an ancilla measurement
			and a reset. \textbf{b,} Phase shift of the logical state extracted
			from the process tomography. During the phase gates, the phases of
			the logical state change linearly with $T_{\mathrm{G}}$. \textbf{c,}
			The gray dotted curve is the process fidelity $F$ of $R_{\mathrm{Kerr}}$ without AQEC and could be regarded as a reference. The AQEC indeed improves
			$F$ except for $T_{G}$ being small. The ET gates
			perform better than $R_{\mathrm{Kerr}}$ as expected. \textbf{d,}
			$F$ in the error and code spaces respectively. In the
			code space, both $R_{\mathrm{Kerr}}$ and $R_{\mathrm{ET}}$ have
			nearly identical $F$. However, in the error space $R_{\mathrm{ET}}$
			has much higher $F$ than $R_{\mathrm{Kerr}}$, corroborating
			that the ET gate is able to protect the state when an error occurs
			during the gate. Note that at $T_{\mathrm{G}}=0$ there is no
			single-photon-loss error yet and the small probability of inferring
			the error space mainly comes from the imperfect AQEC pulse and ancilla
			excitation, consistent with numerical simulations (Supplementary Information).}
		\label{fig:oneQEC}
	\end{figure*}
	
	To demonstrate the ET operations on a bosonic logical qubit, we explore
	a superconducting circuit consisting of a high-quality microwave cavity
	constituting the bosonic logical qubit and a dispersively coupled transmon
	qubit as the ancilla~\cite{wallraff2004,Paik2011}, and the system
	Hamiltonian reads
	\begin{equation}
	H_{0}=\Delta\omega a^{\dagger}a-\chi a^{\dagger}a\left|e\right\rangle \left\langle e\right|-\frac{K}{2}a^{\dagger2}a^{2}.\label{eq:Hamiltonian}
	\end{equation}
	Here, $\Delta\omega$ is the cavity frequency with respect to a carefully
	chosen local oscillator reference, $a^{\dagger}$ ($a$) is the creation
	(annihilation) operator for the bosonic mode, $\ket{e}$ ($\ket{g}$)
	is the excited (ground) state of the ancilla, and $\chi/2\pi=1.60~\mathrm{MHz}$
	and $K/2\pi=4.80~\mathrm{kHz}$ are the dispersive coupling and Kerr
	coefficients originated from the ancilla, respectively. To correct
	the dominant photon-loss errors in the bosonic mode, we encode the
	quantum information on the lowest-order binomial
	code in the cavity~\cite{Michael2016,Hu2019}, which is defined in
	Fock basis as
	\begin{equation}
	|0_{\mathrm{L}}\rangle=\dfrac{|0\rangle+|4\rangle}{\sqrt{2}},\,|1_{\mathrm{L}}\rangle=|2\rangle.\label{eq:logicalState}
	\end{equation}
	We note that proper reference frame needs to be carefully chosen such that there is no accumulation of the relative phase between Fock states $\ket{0}$ and $\ket{4}$. When a single-photon-loss error occurs, the quantum state jumps into
	the error space spanned by the basis states
	\begin{equation}
	|0_{\mathrm{E}}\rangle=|3\rangle,\,|1_{\mathrm{E}}\rangle=|1\rangle.\label{eq:errorState}
	\end{equation}
	When prepare the logical qubit in the code space and set the ancilla
	to the idle state $\left|g\right\rangle $, a phase operation on the
	logical qubit can be easily realized via the Kerr effect since
	\begin{equation}
	\mathcal{P}_{\mathrm{C}}^{\dagger}H_{0}\mathcal{P}_{\mathrm{C}}=K\left(\boldsymbol{I}-\boldsymbol{Z}\right)
	\end{equation}
	with respect to the code basis states (Eq.~\ref{eq:logicalState}).
	Here, $\boldsymbol{I}$ and $\boldsymbol{Z}$ are the Pauli matrices.
	Thus, an arbitrary phase gate $R_{\mathrm{Kerr}}\left(\phi\right)=e^{i\frac{1}{2}\phi\boldsymbol{Z}}$
	on a single logical qubit can be implemented by waiting for a duration
	of $\tau=\phi/2K$. However, such phase operations can not tolerate
	single-photon-loss errors, because the ET condition is not satisfied
	as
	\begin{equation}
	\mathcal{P}_{\mathrm{E}}^{\dagger}H_{0}\mathcal{P}_{\mathrm{E}}=\frac{3}{2}K\boldsymbol{I}.
	\end{equation}
	Furthermore, the cavity's Kerr nonlinearity associated with the coupling
	to the ancilla cannot be switched off, therefore the stored logical qubit
	is always impacted by $R_{\mathrm{Kerr}}$ and suffers random
	photon-loss-error-induced dephasing~\cite{Hu2019}.
	
	To meet the ET condition, we develop a technique to flexibly engineer
	the Hamiltonian in both the code and error spaces. Through a detuned
	microwave drive on the ancilla, photon-number-resolved AC-Stark shift
	(PASS) can be realized. As schematically shown in Fig.~$\,$2\textbf{a},
	due to the strong ancilla-cavity dispersive coupling, the transition
	frequency of the ancilla is photon-number $\left(n\right)$ dependent,
	and thus an off-resonant drive would induce photon-number dependent energy
	shift $\delta_{n}$ due to the AC-Stark effect~\cite{Schuster2005,Gamel2010}.
	Such a frequency shift can also be understood as a geometric phase
	accumulation $\sim\delta_{n}\tau=\frac{\Omega^{2}}{\Delta_{\mathrm{d}}-n\chi}\tau$
	for the joint ancilla-cavity state $\ket{gn}$ (Fig.~$\,$2\textbf{b}), while keeping the excitation to $\ket{en}$ negligible due to the large detuning (Supplementary
	Information). Here $\tau$ is the gate duration time, $\Omega$ is the Rabi drive frequency, and $\Delta_{\mathrm{d}}$ is the drive detuning
	with respect to the ancilla transition frequency corresponding to
	$n=0$. By applying drives with carefully chosen frequencies and amplitudes,
	we could precisely engineer the frequency shifts of the Fock states
	to realize the Hamiltonian
	\begin{equation}
	H_{\mathrm{PASS}}=\sum_{n=0}^{n_{\mathrm{trc}}}\delta_{n}|n\rangle\langle n|,
	\end{equation}
	with the truncated photon number $n_{\mathrm{trc}}=4$ for the code
	considered in this work (Eq.~\ref{eq:logicalState} and Eq.~\ref{eq:errorState}).
	Figure~\ref{fig:wignerFig}\textbf{c} shows the measured Fock state
	frequencies when applying a continuous drive in the middle of the ancilla
	transition frequencies corresponding to $n=3$ and $n=4$, i.e. $\Delta_{\mathrm{d}}=-3.50\chi$
	(the dashed orange line in Fig.~\ref{fig:wignerFig}\textbf{a}).
	The experimental results are well consistent with the theoretical
	predictions.
	
	After experimentally validating the precisely controlled PASS, we
	turn to realize the ET phase gate $R_{\mathrm{ET}}$ on the logical
	qubit. With an appropriate drive amplitude, we can obtain
	\begin{equation}
	\mathcal{P}_{\mathrm{C}}^{\dagger}(H_{0}+H_{\mathrm{PASS}})\mathcal{P}_{\mathrm{C}}=K^{\prime}(\boldsymbol{I}-\boldsymbol{Z})
	\end{equation}
	and
	\begin{equation}
	\mathcal{P}_{\mathrm{E}}^{\dagger}(H_{0}+H_{\mathrm{PASS}})\mathcal{P}_{\mathrm{E}}=K^{\prime}(\boldsymbol{I}-\boldsymbol{Z})+c\boldsymbol{I},
	\end{equation}
	with $K^{\prime}/2\pi=3.33~\mathrm{kHz}$ and $c/2\pi=-0.63~\mathrm{kHz}$.
	Here, the ET condition is satisfied with a re-chosen reference $\Delta\omega=6.09~\mathrm{kHz}$.
	We now verify the ET property of the phase gate by measuring the evolution
	of a logical state $(|0_{L}\rangle-i|1_{L}\rangle)/\sqrt{2}$ with
	($R_{\mathrm{ET}}$) and without ($R_{\mathrm{Kerr}}$) the PASS.
	To separately check the quantum evolution in the code and error spaces,
	we perform Wigner tomography of the output states by post-selecting
	the parity of the excitation number after various evolution times (Fig.~\ref{fig:wignerFig}\textbf{d})
	with the assistance of the ancilla~\cite{SunNature,Hu2019}. The results are
	summarized in Fig.~\ref{fig:wignerFig}\textbf{e}. Comparing $R_{\mathrm{ET}}$
	and $R_{\mathrm{Kerr}}$, the evolution of the logical state in the
	code space shows similar rotations and phase coherence for both cases,
	as indicated by the fringes in the azimuth direction. However, the
	phase coherence in the error space is only preserved by $R_{\mathrm{ET}}$,
	in strong contrast to the significant corruption of phase coherence
	for $R_{\mathrm{Kerr}}$, manifesting the tolerance to the stochastic
	photon-loss error during the ET gate.
	
	Additionally, the ET idle gate $I_{\mathrm{ET}}$ can also be realized
	by simultaneously applying two PASS drives, such that $K^{\prime}=0$.
	This ET realization of an idle Hamiltonian is important, since the
	Kerr effect could be shut off and the quantum information can be protected
	by the binomial QEC code from random photon-loss-error-induced dephasing.
	In Fig.~\ref{fig:wignerFig}\textbf{e}, $I_{\mathrm{ET}}$ shows
	similar ET properties as $R_{\mathrm{ET}}$, while the phase of the
	logical state remains unchanged in both the code and error spaces.
	
	\begin{figure}
		\centering{}\includegraphics{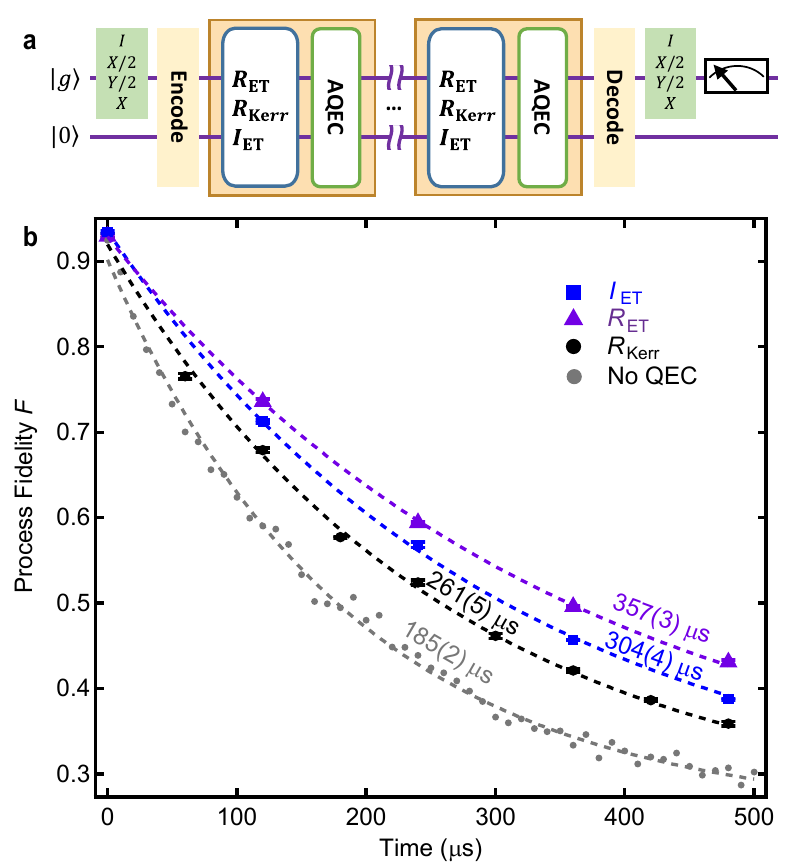} \caption{\textbf{ET gates protected by repetitive AQEC.} \textbf{a,} Experimental
			sequence. \textbf{b,} Experimental process fidelity $F$ as a function
			of time with repetitive and interleaved ET gates and AQECs on the
			logical qubit. Both ET gates have better performance than the non-ET
			$R_{\mathrm{Kerr}}$. The lifetime with the ET idle gate $I_{\mathrm{ET}}$
			is shorter than that with the ET phase gate $R_{\mathrm{ET}}$ because
			of extra ancilla excitation from the additional drive. All these three
			gates with AQEC show better performance than the case without AQEC.}
		\label{fig:QECrep}
	\end{figure}
	
	To demonstrate the potential of the ET gates for fault-tolerant quantum
	computation, we further investigate the ET gates under AQEC protection,
	with the experimental sequence shown in Fig.~3\textbf{a}. An AQEC
	pulse numerically optimized with a duration of $1.5~\mathrm{\mu s}$
	recovers the error state during the ET gate operation and also transfers
	the error entropy associated with the logical state to the ancilla,
	followed by a measurement-based ancilla reset (Methods and Supplementary
	Information). The AQEC is equivalent to previously demonstrated feedback-based
	QEC~\cite{Hu2019}, but holds the advantages of convenience in experiments
	and avoids the latency in the electronic control system since the
	AQEC is error-detection free~\cite{Schindler2011,Reed2012,Waldherr2014}. Figures~\ref{fig:oneQEC}\textbf{b-d}
	summarize the experimental results and process fidelities with different
	gate operation time $T_{\mathrm{G}}$. Arbitrary phase gates can be
	achieved with appropriate $T_{\mathrm{G}}$, however their gate fidelities
	$F$ decay with $T_{\mathrm{G}}$ as expected. In Fig.~\ref{fig:oneQEC}\textbf{c},
	we find all the gates are improved by AQEC when compared with $R_{\mathrm{Kerr}}$
	without AQEC, while the ET gates show superior performances. By measuring
	the process fidelity $F$ in the code and error spaces separately,
	the ET effect is clearly evidenced in Fig.~\ref{fig:oneQEC}\textbf{d}:
	$F$ for $R_{\mathrm{Kerr}}$ and $R_{\mathrm{ET}}$ in the code space
	are almost identical, but $F$ for $R_{\mathrm{ET}}$ in the error
	space is substantially improved.
	
	Finally, the ET logical gate can be interleaved with AQEC and performed
	repeatedly, as illustrated in Fig.~\ref{fig:QECrep}\textbf{a}. Figure~\ref{fig:QECrep}\textbf{b}
	shows the measured process fidelity decaying exponentially as a function
	of time. We have chosen the optimal time interval for each gate ($60~\mu$s
	for $R_{\mathrm{Kerr}}$ and $120~\mu$s for the ET gates). Clearly,
	both ET gates have better performance than the non-ET $R_{\mathrm{Kerr}}$.
	The lifetime with $I_{\mathrm{ET}}$ is shorter than that with $R_{\mathrm{ET}}$
	because of extra ancilla excitation from the additional drive, which
	causes dephasing of the logical state. In addition, all these three gates
	with AQEC have better performance than the case without AQEC, demonstrating
	the effectiveness of AQEC.
	
	%Finally, the ET logical gate can be interleaved with repetitive AQEC, as demonstrated in Fig.~\ref{fig:QECrep}. The results indicate that the repetitive QEC lifetime by calculating the fidelity loss from one additional gate and QEC process. Then we choose the optimal time interval for each gate and perform the repetitive QEC experiment ($60\mu s$ for Kerr gate, $120\mu s$ for ET gate). As shown in Fig.~\ref{fig:QECrep}, the lifetime under ET phase gate is longest, and the ET Idle is noisier because of qubit excitation from the additional drive. The Kerr gate is not ET so it gives a relatively short lifetime.
	
	We introduce the concept of ET gates on a bosonic logical qubit, where
	the evolution in the error space is independent and exactly the same
	as that in the code space. The ET arbitrary phase gates and the idle
	gate have been demonstrated on the lowest-order binomial code by engineering
	the frequency shift of each Fock state, and an enhancement on the
	ET gate fidelity has also been demonstrated with repetitive AQEC.
	Our approach could also be generalized to single-qubit Hadamard gate
	and controlled-phase gate on two binomial logical qubits (Supplementary
	Information), thus constituting the universal ET gate set for quantum
	computation. Therefore, the ET gates and the bosonic QEC codes offer an
	alternative fault-tolerant quantum computation architecture. We note
	that another ET gate on a bosonic logical qubit was independently
	demonstrated in~\cite{Reinhold2019error}, which tolerates the damping
	error of the ancilla by exploiting the ancilla's higher energy levels. These
	two ET gate demonstrations are complementary and together promise
	the ET implementation of QEC against both the damping error of the ancilla
	and photon-loss error of the bosonic mode.
	
	%and two logical qubits in separate 3D cavities have already been realized \cite{Wang2016,Chou2018,Rosenblum2018}
	
	%\noindent \textbf{Online Content}\\ Any methods, additional references, Nature Research reporting summaries, source data, extended data, supplementary information, acknowledgements, peer review information; details of author contributions and competing interests; and statements of data and code availability are available in the online version of the paper.
	
	%\bibliographystyle{Zou}
	%\bibliography{ETgate}
	
	%merlin.mbs apsrev4-1.bst 2010-07-25 4.21a (PWD, AO, DPC) hacked
	%Control: key (0)
	%Control: author (72) initials jnrlst
	%Control: editor formatted (1) identically to author
	%Control: production of article title (0) allowed
	%Control: page (0) single
	%Control: year (1) truncated
	%Control: production of eprint (-1) disabled
	%

	\vbox{}
	
	\noindent \textbf{Acknowledgments}\\This work was supported by National Key Research and Development Program of China (Grant No.2017YFA0304303) and the National Natural Science Foundation of China (Grant No.11474177 and 11874235). C.-L.Z. was supported by National Natural Science Foundation of China (Grant No.11874342 and 11922411) and Anhui Initiative in Quantum Information Technologies (AHY130200).
	
\end{document}

\onecolumngrid
\global\long\def\thefigure{S\arabic{figure}}%
 \setcounter{figure}{0}
\global\long\def\thepage{S\arabic{page}}%
 \setcounter{page}{1}
\global\long\def\theequation{S\arabic{equation}}%
 \setcounter{equation}{0} %\renewcommand{\thesection}{S.\Roman{section}}
\setcounter{section}{0}
\renewcommand{\thetable}{S\arabic{table}}
\setcounter{table}{0}

\title{Supplementary Information for \textquotedblleft Error-transparent
operations on a logical qubit protected by quantum error correction\textquotedblright{}}
\author{Y.~Ma}
\affiliation{Center for Quantum Information, Institute for Interdisciplinary Information
Sciences, Tsinghua University, Beijing 100084, China}
\author{Y.~Xu}
\affiliation{Center for Quantum Information, Institute for Interdisciplinary Information
Sciences, Tsinghua University, Beijing 100084, China}
\author{X.~Mu}
\affiliation{Center for Quantum Information, Institute for Interdisciplinary Information
Sciences, Tsinghua University, Beijing 100084, China}
\author{W.~Cai}
\affiliation{Center for Quantum Information, Institute for Interdisciplinary Information
Sciences, Tsinghua University, Beijing 100084, China}
\author{L.~Hu}
\affiliation{Center for Quantum Information, Institute for Interdisciplinary Information
Sciences, Tsinghua University, Beijing 100084, China}
\author{W.~Wang}
\affiliation{Center for Quantum Information, Institute for Interdisciplinary Information
Sciences, Tsinghua University, Beijing 100084, China}
\author{X.~Pan}
\affiliation{Center for Quantum Information, Institute for Interdisciplinary Information
Sciences, Tsinghua University, Beijing 100084, China}
\author{H.~Wang}
\affiliation{Center for Quantum Information, Institute for Interdisciplinary Information
Sciences, Tsinghua University, Beijing 100084, China}
\author{Y.~P.~Song}
\affiliation{Center for Quantum Information, Institute for Interdisciplinary Information
Sciences, Tsinghua University, Beijing 100084, China}
\author{C.-L.~Zou}
\email{clzou321@ustc.edu.cn}

\affiliation{Key Laboratory of Quantum Information, CAS, University of Science
and Technology of China, Hefei, Anhui 230026, P. R. China}
\author{L.~Sun}
\email{luyansun@tsinghua.edu.cn}

\affiliation{Center for Quantum Information, Institute for Interdisciplinary Information
Sciences, Tsinghua University, Beijing 100084, China}
%\date{\today}
\maketitle

\section{Experimental parameters and techniques}

\subsection{Experimental device}

\begin{figure}[b]
\includegraphics{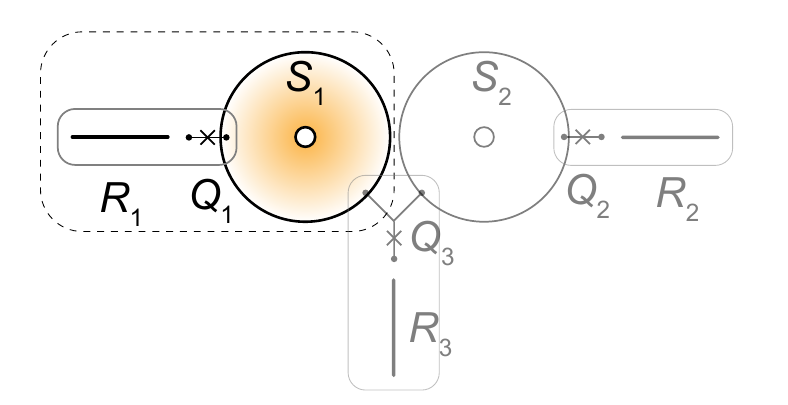}
\caption{\textbf{Experimental device.} The ET gate experiment is based on the parts in the dashed rectangle, consisting of a high-quality three-dimensional coaxial aluminum cavity $S_1$ as the storage cavity to encode the logical qubit, an ancillary transmon qubit $Q_1$, and a stripline cavity $R_1$ for readout of the ancilla. The remaining parts are in their ground states during the experiment.}
\label{fig:ExpDevice}
\end{figure}

The experimental device is composed of two high-quality three-dimensional coaxial aluminum cavities~\cite{Reagor2016,Wang2016Schrodinger,Gao2019} ($S_1$ and $S_2$), three ancillary transmon qubits ($Q_1$, $Q_2$, $Q_3$) and three stripline cavities~\cite{Axline2016} ($R_1$, $R_2$, $R_3$), as shown in Fig.~\ref{fig:ExpDevice}. The detailed geometry of the device can be found in Ref.~\cite{XuGeometricGate}. For current experiment on the error-transparent (ET) gates on a single-logical qubit, we only use the left part of the device ($S_{1}$, $Q_{1}$, and $R_{1}$), and the remaining parts stay in their ground states during the experiment. The single-logical qubit is encoded in the bosonic mode of $S_1$ (referred as the `cavity' henceforth), which is dispersively coupled to the ancilla $Q_1$. The stripline cavity $R_1$ with a high external coupling rate ($\kappa_{out}$) is to readout $Q_1$. The relevant parameters and coherence properties of the device under study are listed in Table~\ref{Table:DeviceParameters} and Table~\ref{Table:coherenttime}. Note that the coefficients $\chi=\chi_{\mathrm{qs}}$ and $K=K_{\mathrm{s}}$ are used in the main manuscript and below for abbreviation.

\begin{table}
\centering %
\begin{tabular}{cc}
\hline
% after \\: \hline or \cline{col1-col2} \cline{col3-col4} ...
Term  & Value\tabularnewline
\hline
$\omega_{\mathrm{q}}/2\pi$  & 6.036 GHz \tabularnewline
$\omega_{\mathrm{s}}/2\pi$  & 6.594 GHz \tabularnewline
$\omega_{\mathrm{r}}/2\pi$  & 8.892 GHz \tabularnewline
\hline
$K_{\mathrm{q}}/2\pi$  & 252 MHz \tabularnewline
$K_{\mathrm{s}}/2\pi$  & 4.8 kHz \tabularnewline
\hline
$\chi_{\mathrm{qs}}/2\pi$  & 1.60 MHz \tabularnewline
$\chi_{\mathrm{qr}}/2\pi$  & 2.00 MHz \tabularnewline
\hline
\end{tabular}
\caption{\textbf{Experimentally characterized parameters for the cavities and the ancilla qubit.} Here, the subscripts (q,s,r) denote the ancilla qubit ($Q_{1}$), storage cavity ($S_{1}$) and readout cavity ($R_{1}$), respectively. $\omega$ is the bare frequency for each compoment, $K$ is the self-Kerr coefficient and $\chi$ is the dispersive coupling strength.}
\label{Table:DeviceParameters}
\end{table}

\begin{table}
\centering %
\begin{tabular}{ccccc}
\hline
% after \\: \hline or \cline{col1-col2} \cline{col3-col4} ...
$\quad$  & $Q_{1}$  & $Q_{3}$  & $S_{1}$  & $R_{1}$ \tabularnewline
\hline
$T_{1}$  & $35\ \mu$s  & $25\ \mu$s  & $480\ \mu$s  & $58$ ns \tabularnewline
$T_{2}$  & $25\ \mu$s  & $30\ \mu$s  & $560\ \mu$s  & - \tabularnewline
$T_{\phi}$  & $39\ \mu$s  & $75\ \mu$s  & 1.3 ms  & - \tabularnewline
\hline
$n_{\mathrm{th}}$  & $1.6\%$  & $0.7\%$  & $<1\%$  & - \tabularnewline
\hline
\end{tabular}
\caption{\textbf{Coherence properties of the ancilla qubits and the cavities.} $T_{1}$ and $T_{2}$ are the experimentally measured energy and phase relaxation times, respectively, $T_{\phi}$ is the derived pure dephasing time, and $n_{\mathrm{th}}$ is the thermal excitation in the experiment. }
\label{Table:coherenttime}
\end{table}

\subsection{Experimental techniques}

In this work, we have developed two experimental techniques to realize
the ET gates under the protection of repetitive quantum error correction (QEC). The first technique is the photon-number-resolved AC-Stark shift (PASS) to engineer the system Hamiltonian precisely, and thus the ET condition can be satisfied for the binomial codes. The second technique is the autonomous QEC (AQEC), by which the single-photon-loss error can be detected and corrected without extracting the error syndrome by the control electronics and thus the electronic latency can be avoided. In this section, we provide the details of the two techniques.

\subsubsection{PASS}

For the system under study, the Hamiltonian for realizing PASS reads
\begin{eqnarray}
H_{0} & = & \Delta\omega a^{\dagger}a-\frac{K}{2}a^{\dagger2}a^{2}+\left(\omega_{\mathrm{q}}-\chi a^{\dagger}a-\omega_{\mathrm{d}}\right)\left|e\right\rangle \left\langle e\right|\nonumber \\
 &  & +\Omega\left(\left|e\right\rangle \left\langle g\right|+\left|g\right\rangle \left\langle e\right|\right),
\end{eqnarray}
where $\Delta\omega$ is the cavity frequency with respect to a carefully chosen local oscillator reference, $a^{\dagger}$ ($a$) is the creation (annihilation) operator for the bosonic mode, $\ket{e}$ ($\ket{g}$) is the excited (ground) state of the ancilla, $\chi$ is the dispersive coupling strength, $K$ is the self-Kerr coefficient of the cavity originated from the ancilla, $\omega_{\mathrm{q}}$ is the ancilla qubit frequency when the cavity is in vacuum, $\omega_{\mathrm{d}}$ is the driving frequency on the ancilla, and $\Omega$ is the Rabi drive frequency. In the following, we define $\Delta_{\mathrm{d}}\equiv\omega_{\mathrm{q}}-\omega_{\mathrm{d}}$.
In the limit of $\Omega\ll\left|\Delta_{\mathrm{d}}-\chi a^{\dagger}a\right|$
for all cavity states, we have the effective Hamiltonian as \citep{Gamel2010}
\begin{eqnarray}
H_{\mathrm{eff}} & \approx & \Delta\omega a^{\dagger}a-\frac{K}{2}a^{\dagger2}a^{2}+\left(\Delta_{\mathrm{d}}-\chi a^{\dagger}a\right)\left|e\right\rangle \left\langle e\right|\nonumber \\
 &  & +\frac{\Omega^{2}}{\left(\Delta_{\mathrm{d}}-\chi a^{\dagger}a\right)}\left(\left|e\right\rangle \left\langle e\right|-\left|g\right\rangle \left\langle g\right|\right).
\end{eqnarray}
Then, the effective frequencies for $\left|gn\right\rangle $
and $\left|en\right\rangle $ (the joint ancilla-cavity states) are $n\Delta\omega-n\left(n-1\right)\frac{K}{2}-\frac{\Omega^{2}}{\left(\Delta_{\mathrm{d}}-n\chi \right)}$
and $n\Delta\omega-n\left(n-1\right)\frac{K}{2}+\left(\Delta_{\mathrm{d}}-n\chi\right)+\frac{\Omega^{2}}{\left(\Delta_{\mathrm{d}}-n\chi \right)}$,
respectively. Therefore, the effective frequency for Fock state $\ket{n}$ is
shifted by $\pm\frac{\Omega^{2}}{\left(\Delta_{\mathrm{d}}-n\chi\right)}$
conditional on $n$ and the state of the ancilla qubit. As a result,
the detuned drive induces the PASS.
For example, for the drive frequency lying between ancilla transition frequencies corresponding to $\left|n\right\rangle $
and $\left|n+1\right\rangle $, i.e. $\omega_{\mathrm{q}}-\left(n+1\right)\chi<\omega_{\mathrm{d}}<\omega_{\mathrm{q}}-n\chi $, the PASS on $\left|gn\right\rangle $ and $\left|g\left(n+1\right)\right\rangle $
are negative and positive, respectively.

\begin{figure}
\includegraphics{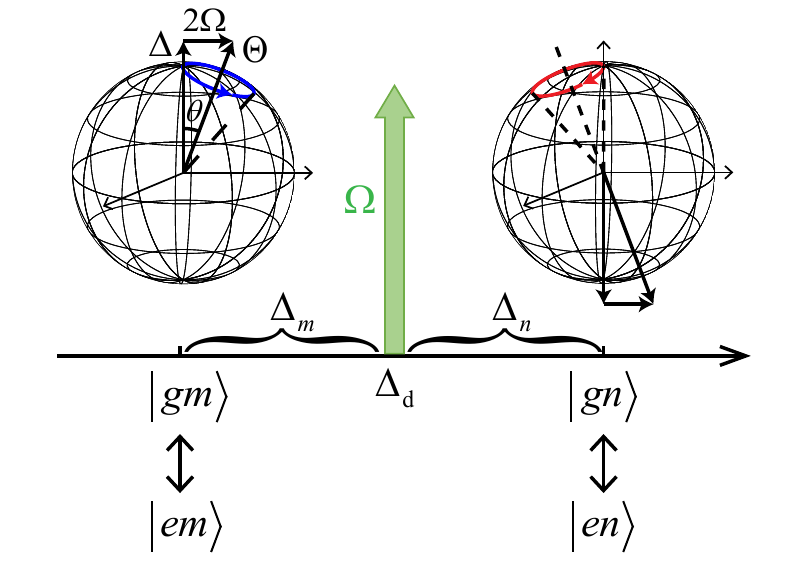} \caption{\textbf{Geometric phase interpretation of the PASS.} An off-resonant drive with a frequency $\Delta_{\mathrm{d}}$ between two dispersive transition frequencies of the ancilla and an amplitude $\Omega$ would produce a geometric phase on each ancilla state. This phase is accumulated constantly and causes an equivalent frequency shift on the photon Fock state because the ancilla is approximately in its ground state and can be traced out. The direction of the frequency shift is related to the sign of the detuning $\Delta$.}
\label{fig:OffRes}
\end{figure}

The PASS can also be understood from the point view of the geometric phase
accumulated on the photonic Fock state due to the off-resonant
drive, as shown in Fig.~\ref{fig:OffRes}. The ancilla state initialized at the pole cannot be efficiently excited, but only rotates near the pole. The solid angle enclosed by the trajectory of the state on
the Bloch sphere can be represented as
\begin{equation}
\Theta=2\pi[1-\cos(\theta)]=2\pi(1-\frac{\Delta}{\sqrt{4\Omega^{2}+\Delta^{2}}}).
\label{eq:solidAngle}
\end{equation}
For the detuned drive, the cycle period for the state rotating on
the Bloch sphere is
\begin{equation}
T_{\mathrm{cyc}}=\frac{2\pi}{\sqrt{4\Omega^{2}+\Delta^{2}}},\label{eq:cycleTime}
\end{equation}
so the accumulated geometric phase ($\Theta/2$) is proportional to the
number of rotation cycles, and the equivalent frequency shift can
be derived as
\begin{align}
\delta(\epsilon,\Delta) & =\frac{\Theta}{2T_{\mathrm{cyc}}}=\frac{\sqrt{4\Omega^{2}+\Delta^{2}}-\Delta}{2}\label{eq:freqShift}\\
 & \approx\frac{\Omega^{2}}{\Delta}
\end{align}
for $\left|\Delta/\Omega\right|\gg1$.

By taking into account the amplitude decay rate of the excited state of the ancilla
qubit ($\kappa_{\mathrm{q}}$), and for the ancilla prepared
in the ground state, the effective Hamiltonian becomes
\begin{eqnarray}
H_{\mathrm{eff}} & \approx & \Delta\omega a^{\dagger}a-\frac{K}{2}a^{\dagger2}a^{2}-\frac{\Omega^{2}}{\left(\Delta_{\mathrm{d}}-\chi a^{\dagger}a\right)-i\kappa_{\mathrm{q}}}.
\end{eqnarray}
Therefore, the PASS drive would not only induce the energy level shift
\begin{equation}
H_{\mathrm{PASS}}=\sum_{n}\delta_{n}|n\rangle\langle n|=\sum_{n}\frac{\Omega^{2}}{\Delta_{\mathrm{d}}-n\chi}|n\rangle\langle n|,
\label{eq:PASS}
\end{equation}
%with $\delta_{n}=\frac{\Omega^{2}}{\Delta_{\mathrm{d}}-n\chi}$,
but also induce the phase decoherence of the Fock states with a rate
\begin{equation}
\gamma_{n}\approx\frac{\Omega^{2}}{\left(\Delta_{\mathrm{d}}-n\chi\right)^{2}}\kappa_{\mathrm{q}}.\label{eq:induced-decoherence}
\end{equation}
This equation indicates that the frequency shift can be used to
implement a logical-qubit phase gate while the error brought from the
ancilla is significantly suppressed because of the ancilla's small excitation
during the gate.

Here, we also want to briefly discuss the limitation of the PASS technique.
For the purpose of error transparency in this work, the PASS should
compensate the self-Kerr effect, i.e. $\delta_{n}=\mathcal{O}\left(K\right)$.
If an individual PASS drive is applied to $\ket{gn}$ and $\ket{g(n+1)}$, we would have $\Delta_{\mathrm{d}}-n\chi\approx\frac{\chi}{2}$, and
thus the compensation requires $\frac{2\Omega^{2}}{\chi}=\mathcal{O}\left(K\right)$.
In addition, the PASS requires a small drive amplitude $\Omega\ll\chi$
and negligible induced decoherence $\gamma_{n}\ll n\kappa_{\mathrm{a}}$,
with $\kappa_{\mathrm{a}}$ being the amplitude decay rate of the cavity.
From Eq.~(\ref{eq:induced-decoherence}), we have $\gamma_{n}\approx\frac{4\Omega^{2}}{\chi}\frac{\kappa_{\mathrm{q}}}{\chi}=\frac{2\kappa_{\mathrm{q}}}{\chi}\mathcal{O}\left(K\right)$.
Taking the fact that the self-Kerr coefficient of the cavity is related
to the cross-Kerr coefficient (dispersive coupling strength) as $K=\chi^{2}/4E_{c}$, with $E_{c}$ being the anharmonicity (the self-Kerr) of the ancillary transmon qubit, we can derive the conditions for the PASS drive to achieve the ET gates as: (1) $\Omega=\frac{\chi}{2}\mathcal{O}\left(\sqrt{\frac{\chi}{2E_{c}}}\right)$,
(2) $\frac{1}{2}\sqrt{\frac{\chi}{2E_{c}}}\ll1$, and (3) $\frac{\chi}{2E_{c}}\ll\frac{\kappa_{\mathrm{a}}}{\kappa_{\mathrm{q}}}$.
For the device in this study, we have $E_{c}/2\pi\sim252~\mathrm{MHz}$,
$\chi/2\pi\sim1.60~\mathrm{MHz}$ and $\frac{\kappa_{\mathrm{a}}}{\kappa_{\mathrm{q}}}\sim0.07$,
therefore $\frac{\chi}{E_{c}}\sim6.4\times10^{-3}$ and all the above conditions
are satisfied. For a better performance of the PASS technique for ET gates,
$\kappa_{\mathrm{q}}$ and $\frac{\chi}{E_{c}}$ of the superconducting
circuit should be further reduced.

\subsubsection{AQEC}

AQEC is equivalent to the standard measurement-based QEC, which consists of both error detection and correction operations. However, AQEC does not need error detections. It is worth noting that the AQEC had been used with the three-qubit repetition code~\cite{Schindler2011,Reed2012,Waldherr2014}, while its extension to the bosonic codes requires rather sophisticated conditional unitary operations. To perform AQEC for the bosonic codes, a unitary transition is implemented to correct the logical state in the error space $\ket{\psi_{\mathrm{E}}}$ while driving the ancilla to an orthogonal state:
\begin{equation}
U|\psi_{\mathrm{E}}\rangle|g\rangle=|\psi_{\mathrm{L}}\rangle|e\rangle,\label{eq:defAQEC}
\end{equation}
but keep the logical state in the code space $\ket{\psi_{\mathrm{L}}}$ unchanged:
\begin{equation}
U|\psi_{\mathrm{L}}\rangle|g\rangle=|\psi_{\mathrm{L}}\rangle|g\rangle.
\end{equation}
In this case, the correlation between the quantum system and the environment (which
induces errors) is erased, and thus the error entropy is transferred to the ancilla. After the operation, the logical state is recovered, and the
ancilla system can be traced out. Note that in our experiment $\ket{\psi_{\mathrm{E}}}=a\ket{\psi_{\mathrm{L}}}$ (single-photon-loss error) in the error space; while there is a non-unitary no-jump evolution $e^{-\kappa_{\mathrm{a}}a^{\dagger}a t/2}$ of $\ket{\psi_{\mathrm{L}}}$ in the code space, which is corrected in the AQEC pulse.

In practice, we need to reuse the ancilla, therefore we reset the ancilla
to a pure state after each implementation of AQEC for the next round of AQEC. Compared with previous demonstrations of QEC~\citep{Hu2019}, the whole process does not need any projective measurement on the encoded bosonic state, and the error syndrome is not necessarily to be extracted. Therefore,
the real-time feedback control system is not required any more and the potential electronic latency is avoided. The reset of the ancilla could be implemented in either digital or analog approaches. For the digital approach, the ancilla can be directly readout, and a control pulse dependent on the readout result is then applied to reset the ancilla. In this measurement-based case, the digital control only needs to implement the reset before the next AQEC step, which is hundreds of microseconds later in our case, in contrast to a few hundred nanoseconds required for real-time feedback control. For the analog approach, the ancilla could be engineered to couple to a readout cavity by switching on a stimulating drive, which can allow the decay of the excitation in the ancilla within about one microsecond. Here in this experiment, we use the so-called gradient ascent pulse engineering (GRAPE) algorithm~\cite{Khaneja2005,deFouquieres2011} to numerically optimize the AQEC pulse and use measurement-feedback method to reset the ancilla.

\section{More Experimental data}

\subsection{Detailed experimental data of the PASS}

\begin{table*}[t]
\vspace{12bp}

\begin{centering}
\begin{tabular*}{0.9\textwidth}{@{\extracolsep{\fill}}@{\extracolsep{\fill}}@{\extracolsep{\fill}}ccc}
\hline
% after \\: \hline or \cline{col1-col2} \cline{col3-col4} ...
Driving parameter & ET phase gate & ET Idle\tabularnewline
\hline
Amplitude $\Omega$ & $0.074\chi$ & $0.054\chi,\,0.074\chi$\tabularnewline
Detuning $\Delta_{\mathrm{d}}$ & $-3.41\chi$ & $-3.37\chi,\,-2.27\chi$\tabularnewline
\hline
\end{tabular*}
\par\end{centering}
\vspace{12bp}

\begin{centering}
\begin{tabular*}{0.9\textwidth}{@{\extracolsep{\fill}}@{\extracolsep{\fill}}@{\extracolsep{\fill}}cccccccc}
\hline
% after \\: \hline or \cline{col1-col2} \cline{col3-col4} ...
Fock state frequency shift ($2\pi\,$kHz) & $|1\rangle$ & $|2\rangle$ & $|3\rangle$ & $|4\rangle$ & $f_{3}-f_{1}$ & $f_{4}-f_{2}$ & $f_{4}/2-f_{2}$\tabularnewline
\hline
phase gate due to Kerr $R_{\mathrm{Kerr}}$ & $0.14(3)$ & $-4.55(3)$ & $-14.39(3)$ & $-28.92(6)$ & $-14.25$ & $-24.37$ & $-9.91$\tabularnewline
ET phase gate $R_{\mathrm{ET}}$ & $-0.10(2)$ & $-5.52(1)$ & $-18.90(4)$ & $-24.37(7)$ & $-18.80$ & $-18.85$ & $-6.67$\tabularnewline
ET idle gate $I_{\mathrm{ET}}$ & $-0.88(3)$ & $-12.16(3)$ & $-13.05(3)$ & $-24.34(7)$ & $-12.17$ & $-12.18$ & $-0.02$\tabularnewline
\hline
\end{tabular*}
\par\end{centering}
\caption{\textbf{Parameters of the PASS drives and frequency shifts in the experiment.} The numbers in the parenthesis are the measurement uncertainty.}
\label{T:expPara}
\end{table*}

\begin{figure}
\includegraphics{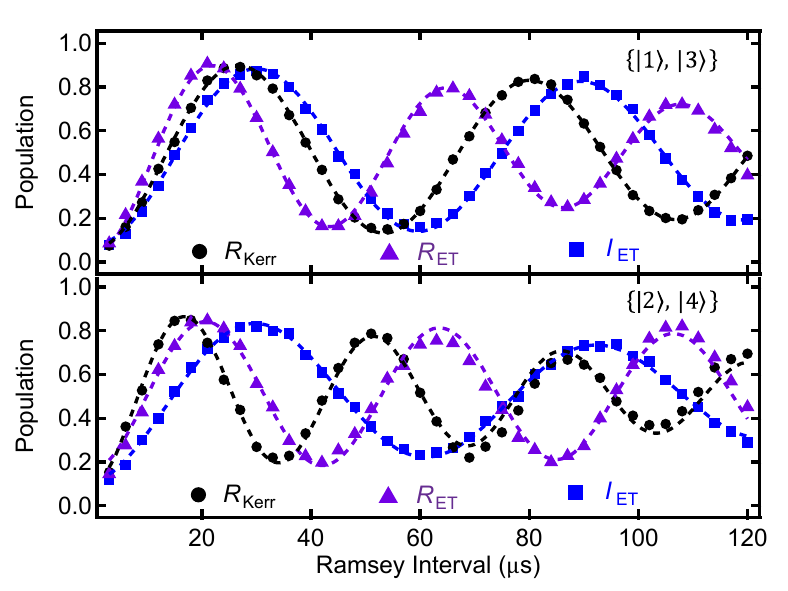}
\caption{\textbf{Ramsey-type experiments in the code and error spaces.} The Ramsey experiment is performed with a superposition of Fock states $|1\rangle$ and $|3\rangle$ (the error space basis); and Fock states $|2\rangle$ and $|4\rangle$ (the code space basis). The synchronized oscillations with both one and two PASS drives in the two spaces demonstrate the satisfaction of the ET condition. However, when the PASS drive is off, the oscillations are not synchronized any more, indicating the non-ET condition. The dashed lines are fits with a decayed sinusoidal function.}
\label{fig:suppleETcon}
\end{figure}

\begin{figure*}
\includegraphics{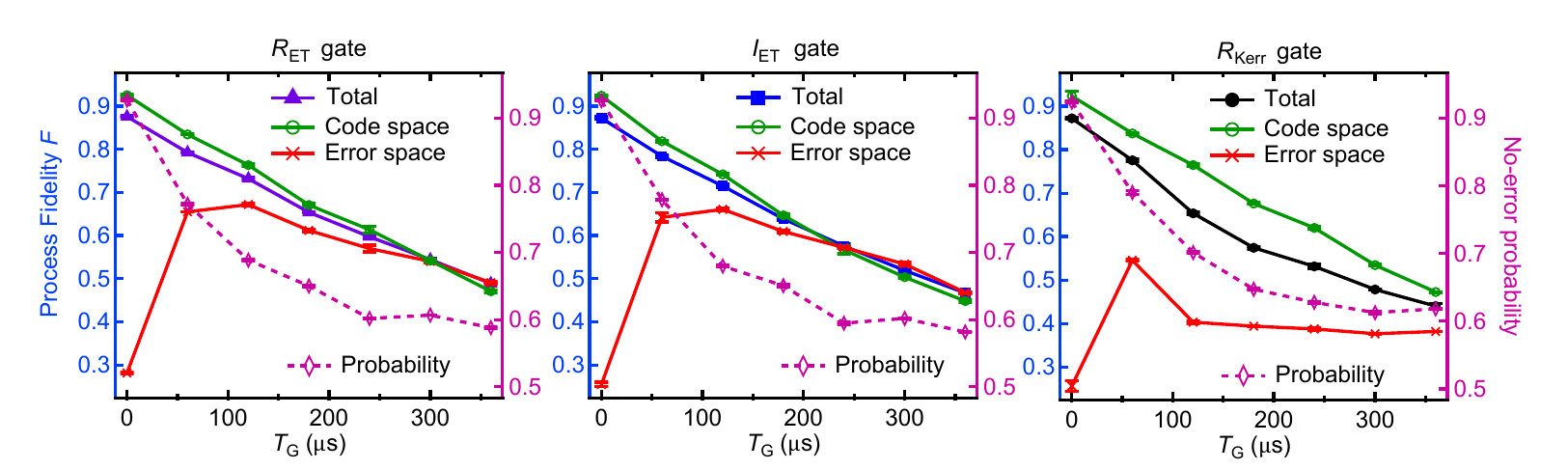} \caption{\textbf{Process fidelity of the three single-logical-qubit gates with AQEC in the error space and code space respectively.} The dashed lines corresponding to the right vertical axis are the probabilities of detecting no-error (no single-photon loss).}
\label{fig:suppleExpData}
\end{figure*}

In the experiment, we have used one or two microwave drives to precisely control the energy shifts of the photon number states, according to the photon-number-resolved AC-Stark effect. The experimental technique and theoretical details are provided in the next section. Here, we show extra experimental results  complementary to the presented ones in the main text.

Table~\ref{T:expPara} summarizes the parameters of the microwave drives used in the experiment. $\Delta_{\mathrm{d}}$ is the
drive detuning with respect to the ancilla qubit transition frequency when the cavity is in a vacuum state (i.e. there is no photon-induced frequency shift to the ancilla qubit), and $\Omega$ is the corresponding Rabi drive frequency that is proportional to the microwave driving amplitude. Here, all the driving parameters are optimized to minimize the excitation of the ancilla qubit in the simulation, and then carefully calibrated in the experiment.

For the ET phase gate, there is only one microwave drive with $\Delta_{\mathrm{d}}$ being between $-3\chi$ and $-4\chi$. Therefore, this drive makes the frequency shifts of Fock states $|3\rangle$ and $|4\rangle$
in opposite directions, thus satisfying the ET condition $(f_{4}-f_{2})-(f_{3}-f_{1})=0$. Here $f_n$ are measured through Ramsey-type experiments on superposition states $(\ket{0}+\ket{n})/\sqrt{2}$ with $\Delta \omega=0$. The experimentally measured $f_n$ for the three different gates $R_{\mathrm{Kerr}}$, $R_{\mathrm{ET}}$, and $I_{\mathrm{ET}}$ are also provided in Table~\ref{T:expPara}. We find that the ET condition is indeed satisfied. In comparison, the case without the PASS drive (the phase gate $R_{\mathrm{Kerr}}$ due to the Kerr coefficient) has $(f_{4}-f_{2})-(f_{3}-f_{1})\approx2\pi\times10~\mathrm{kHz}$. For the ET gate $R_{\mathrm{ET}}$, we choose a cavity frequency in a reference frame with $\Delta\omega/2\pi=6.09~\mathrm{kHz}$ to fix the phase of Fock state $|4\rangle$, and then there is a non-zero rotating frequency of Fock state $|2\rangle$ relative to Fock states $|0\rangle$ and $|4\rangle$, i.e. $f_{4}/2-f_{2}=6.67$~kHz. As a result, a phase gate can be realized to the binomial code.

By adding one more PASS drive with $\Delta_{\mathrm{d}}$ being between $-2\chi$ and $-3\chi$, we can further compensate the relative rotating frequency between $\ket{0_{\mathrm{L}}}=(\ket{0}+\ket{4})/\sqrt{2}$ and $\ket{1_{\mathrm{L}}}=\ket{2}$, and essentially generate the ET idle gate $I_{\mathrm{ET}}$ in the same reference frame as $R_{\mathrm{ET}}$. With the additional PASS drive, the induced frequency shifts of Fock states $|2\rangle$ and $|3\rangle$ have different directions. From Table~\ref{T:expPara}, we find that $(f_{4}-f_{2})-(f_{3}-f_{1})=0$ and $f_{4}/2-f_{2}=0$ are both satisfied, indicating the satisfaction of both the ET condition and the idle operation condition. We note that such a gate is important for our experimental system, because the idle gate could be used to protect the quantum information together with QEC and the corruption of quantum information due to the Kerr effect can be eliminated.

The above ET/non-ET conditions with/without the PASS drives have been experimentally verified in Ramsey-type experiments, as shown in Fig.~\ref{fig:suppleETcon}. The synchronized oscillations with one or two PASS drives in the code and error spaces demonstrate the satisfaction of the ET condition. However, when the PASS drive is off, the oscillations are not synchronized any more, indicating the non-ET condition.

\begin{figure*}
\includegraphics{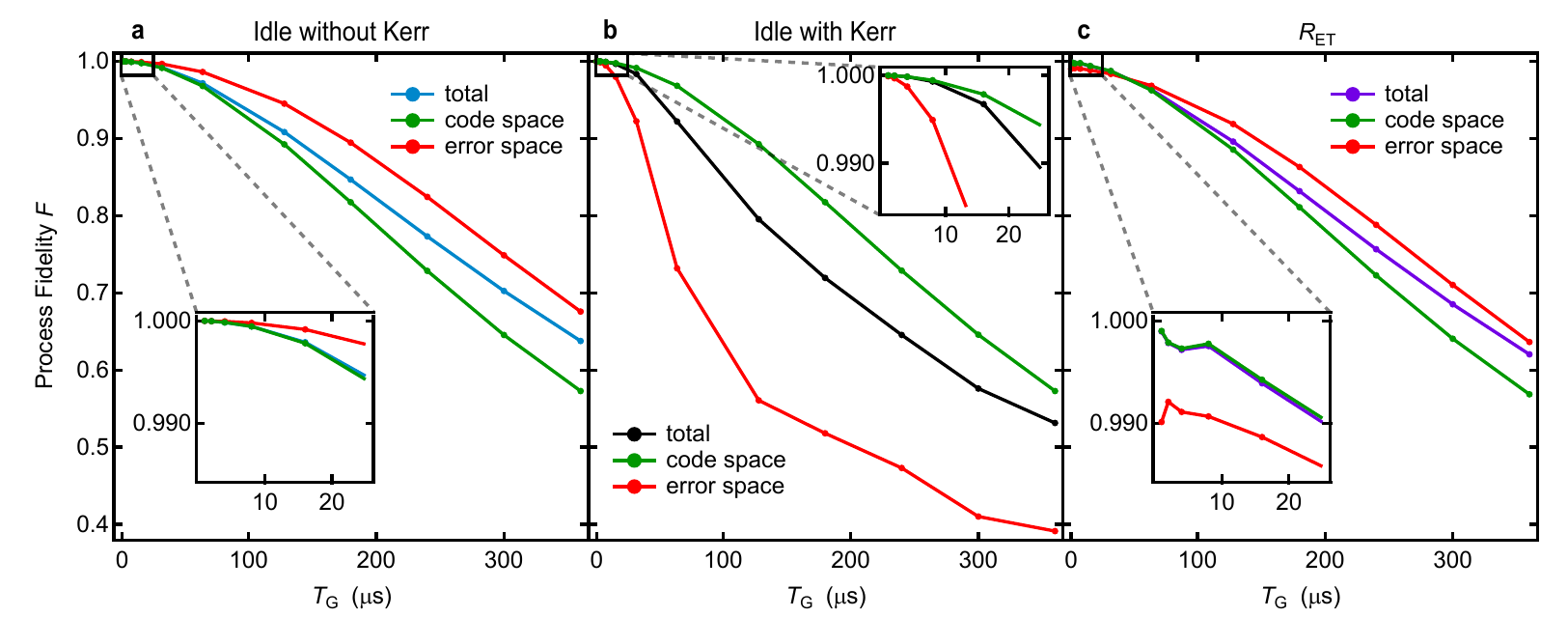} \caption{\textbf{Numerical simulation results with ideal AQEC.} \textbf{a,} The ideal idle gate with no cavity's self-Kerr effect. \textbf{b,} The idle gate with cavity's self-Kerr effect. The logical state in the code space can still preserve, but corrupts quickly in the error space because the ET condition is not satisfied. \textbf{c,} ET phase gate with the PASS drive. The main trend is the same as \textbf{a} when $T_{G}$ is sufficiently large. However, there is a little loss when $T_{G}$ approaches zero because the effective Hamiltonian is no longer satisfied in this regime.}
\label{fig:IdealSimu}
\end{figure*}

\begin{figure*}
\includegraphics{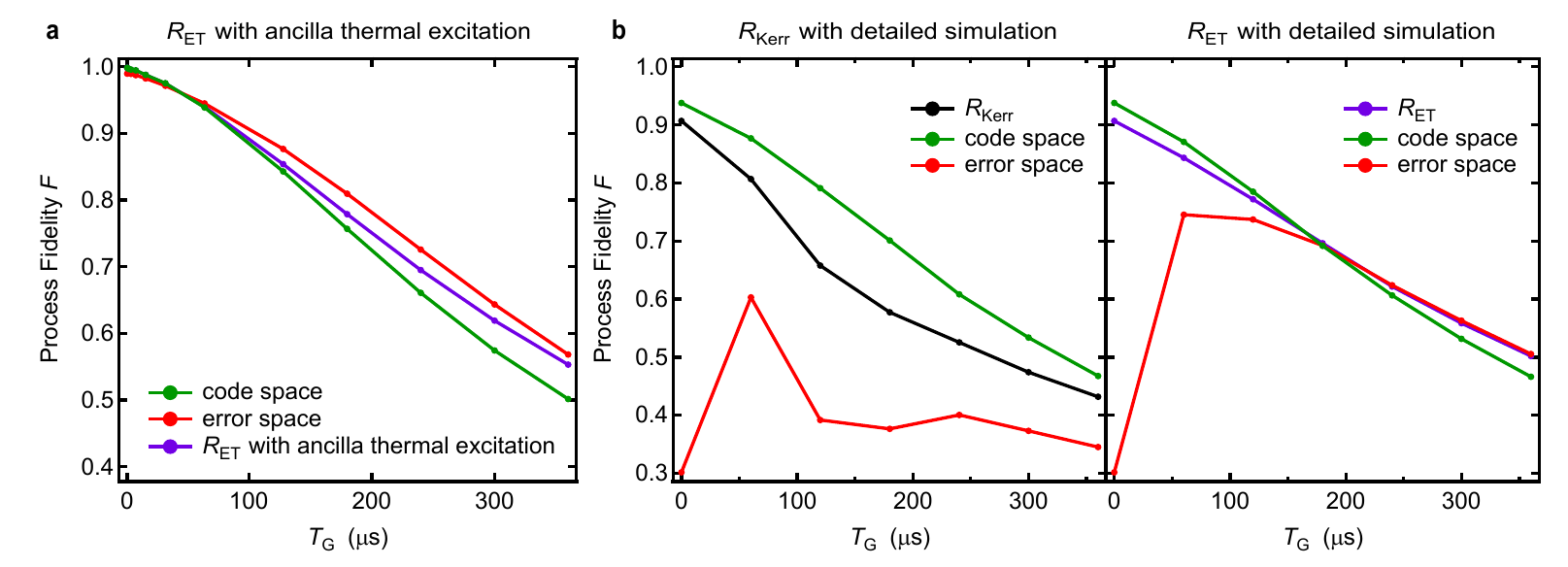} \caption{\textbf{Numerical simulation results with ancilla thermal excitation and operation errors.} \textbf{a,} The ET phase gate with the ancilla thermal excitation, but still with ideal AQEC. The fidelity decays more linearly, in contrast to the quadratically decay curves in Fig.~\ref{fig:IdealSimu}\textbf{c} with zero bath temperature. \textbf{b,} More detailed simulation of the non-ET phase gate $R_{\mathrm{Kerr}}$ and the ET phase gate $R_{\mathrm{ET}}$ including all possible imperfections. The results in \textbf{b} are very similar to the measured ones in Fig.~3\textbf{c} of the main text.}
\label{fig:DetailedSimu}
\end{figure*}

\subsection{Process fidelities for the single-logical-qubit gates with AQEC}

Figure~\ref{fig:suppleExpData} provides more concrete data for the three different gates, accompanying Fig.$\,$3 in the main text. The performances of the ET and non-ET gates are characterized by measuring the process fidelity $F$ as a function of the gate time $T_{\mathrm{G}}$. Here, $F$ is separately measured for the code and error spaces by post-selecting the ancilla state that indicates if an error happens or not. The experimentally measured probabilities of no-error happening are also provided, by which the total fidelity can be derived as a weighted combination of those in both the error and code spaces.

The main feature for the three gates in Figure~\ref{fig:suppleExpData}
is that the fidelities in the error space for the ET gates $R_{\mathrm{ET}}$
and $I_{\mathrm{ET}}$ are very similar and much higher than the non-ET gate $R_{\mathrm{Kerr}}$. Such a difference
manifests that the ET gates possess the capability of protecting the quantum
information in the error space from corruption.

The no-error probability decays with $T_{\mathrm{G}}$ as expected, and so does the fidelity in the code space for all these three gates. However, the fidelities in the error space are low for $T_{\mathrm{G}}=0$, and then jump to a peak value followed by a decay. The reason for this unexpected behavior with small $T_{\mathrm{G}}$ is that in these cases the error is mainly induced by the ancilla decoherence, ancilla excitation, and operation errors, instead of the single-photon-loss error (has not happened yet) that the ET gates can protect. When $T_{G}$ is large enough, the fidelity in the error space becomes much higher because the contribution of single-photon-loss errors dominates in the error space. When $T_{\mathrm{G}}$ further increases, the uncorrectable high-order
photon-loss errors happen with higher probabilities, and therefore the overall
fidelities in both the code and error spaces decay.

For the non-ET gate, the fidelity decays much faster because of the fast dephasing of quantum information in the error space due to the non-ET Kerr effect. Although
the non-ET gate loses the phase information in the error space quickly, its fidelity remains at about $0.4$ since the probability distribution of the basis states preserves, in good agreement with numerical simulations (see the numerical analysis below). It is also worth noting that the fidelities in the code space are slightly lower under the ET gates because of the additional ancilla excitation caused by the off-resonant drive (see the theory section for more discussions).

\section{Numerical analysis}

To analyze sources of errors in the experiment and study the viability
of the ET gates for potential fault-tolerant quantum computation, we implement numerical simulations according to our experimentally calibrated parameters. It is anticipated that the main experimental imperfections include: (1) the imperfect AQEC operations on the system, (2) the decoherence of the ancilla qubit during the ET gates, and (3) the thermal excitation of the ancilla qubit. In the following, we compare the results under different situations with the noiseless ideal case.

\subsection{Numerical simulation with ideal AQEC}

First of all, we use the ideal AQEC process in the numerical simulation
to study the viability of the PASS technique for the realization of ET phase gates. The decoherence (damping and dephasing) of both the ancilla and the cavity are included in the numerical model. However, we assume the thermal bath are in
the vacuum state, so the ancilla cannot be populated by the thermal noise
from the bath. By substituting the experimentally calibrated parameters into
the master equations, the system evolutions are numerically solved and the corresponding process fidelities are summarized in Fig.~\ref{fig:IdealSimu}.
Similar to the experiments in the main text, the ancilla qubit is
traced out after the ET gate evolution with different gate times, and the
density matrices in the error and code spaces are separately processed
to obtain the process fidelities.

Comparing the results from the ET and non-ET gates in Fig.~\ref{fig:IdealSimu},
we find the performances of the ET gates (Fig.~\ref{fig:IdealSimu}\textbf{c}) are close to the ideal case without the cavity's self-Kerr effect (Fig.~\ref{fig:IdealSimu}\textbf{a}). No self-Kerr effect of the cavity is a pre-assumption for many theoretical works on quantum gates~\cite{Albert2018}. However, when the self-Kerr presents in the cavity, the idle operation does not satisfy the ET condition, so the state in the error space corrupts quickly (Fig.~\ref{fig:IdealSimu}\textbf{b}).

By applying the PASS technique on the system to engineer the Hamiltonian,
the ET condition can be satisfied and the process fidelity in the error space preserves. However, we find that the curves of the ET phase gates are slightly lower than the ideal case. The reason could be attributed to the excitation of the ancilla by the off-resonant drive. As predicted by Eq.~\ref{eq:induced-decoherence}, the state jump of the ancilla qubit would induce the dephasing of the cavity states. In addition, there is a small loss of fidelity when $T_{G}$ approaches zero (Fig~\ref{fig:IdealSimu}\textbf{c} inset). This is because the effective Hamiltonian approximation for PASS is only satisfied when the gate time is sufficiently long $T_{\mathrm{G}}\gg1/\Delta$. Therefore, for $T_{\mathrm{G}}<10~\mathrm{\mu s}$ in our experiment, the geometric phase cannot be regarded as continuously accumulated, and the PASS is deviated from Eq.~\ref{eq:PASS}.

\subsection{Numerical simulation with ancilla thermal excitation and imperfect AQEC}

The above simulations reveal the effect of the imperfections due to the drive-induced ancilla excitation and the consequent ancilla-excitation-induced
dephasing, as predicted in the previous section on PASS.
In practice, there are more imperfections that could induce the loss
of gate fidelity.

\begin{figure}
\includegraphics{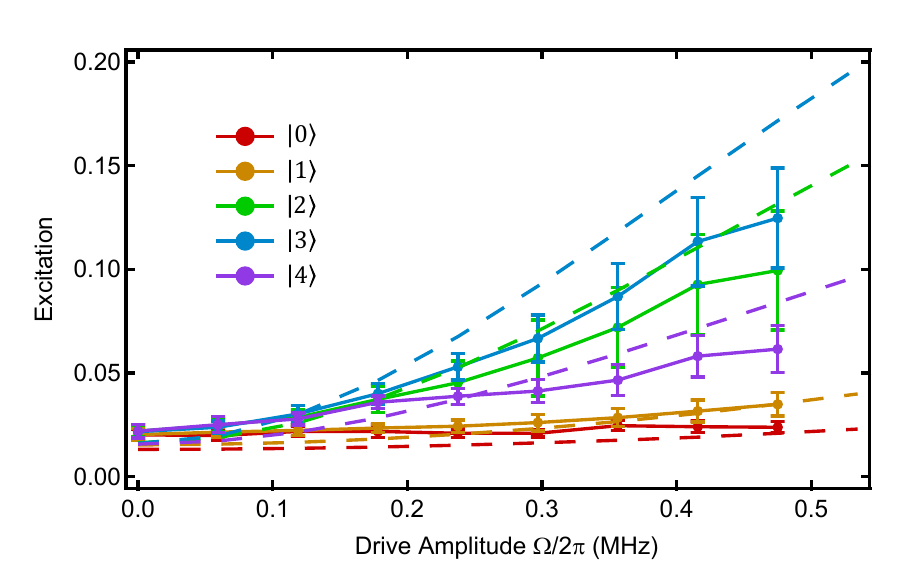} \caption{\textbf{Ancilla excitation from the PASS drive.} The ancilla excitation due to an off-resonant drive with a detuning frequency $\Delta_{\mathrm{d}}=-2.5\chi$ are measured as a function of the drive amplitude when the cavity is initialized at differnet Fock states. Dotted lines are simulation results. The ET gate used in the experiment is performed with the drive amplitude $\Omega/2\pi \approx 0.1~$MHz. Therefore, the additional excitation due to the PASS drive is about $0.01$, while the thermal excitation is about $0.02$.}
\label{fig:OffResExcit}
\end{figure}

One main contribution to the loss is the ancilla thermal
excitation. The results including the ancilla thermal bath are shown in Fig.~\ref{fig:DetailedSimu}\textbf{a}. The fidelity decays more linearly,
in contrast to the quadratically decay curves in Fig.~\ref{fig:IdealSimu}\textbf{c}
with zero bath temperature. As shown in Fig.~\ref{fig:OffResExcit},
the ancilla excitation caused by the PASS drive increases with the drive amplitude.
For the parameter used in our ET experiments ($\Omega/2\pi\sim0.1\,\mathrm{MHz}$),
the PASS-drive-induced excitation is smaller than the measured thermal excitation in the experiment. Therefore, we conclude that the photon dephasing
is mainly from the ancilla thermal excitation, which can be suppressed by a cold bath. The PASS-drive-induced dephasing could be potentially solved by using alternative bosonic codes that are robust to dephasing errors, such as the cat code~\cite{Leghtas2013,Mirrahimi2014} or the numerically optimized codes~\cite{Michael2016,Li2017,Li2019}.

In the experiment, there are inevitable imperfections in the AQEC pulse and the measurement-feedback operation. To account for these errors on the performance of the non-ET gate $R_{\mathrm{Kerr}}$ and the ET phase gate $R_{\mathrm{ET}}$, a more detailed simulation including these imperfections are
carried out, and the results are shown in Fig.~\ref{fig:DetailedSimu}\textbf{b}.
The obtained results agree well with the experimental results in Fig.~3 of the main text, indicating the main imperfections in our experiments are due to the thermal excitation of the ancilla qubit and the operation errors.

\section{Theory}

\subsection{Requirement for error transparency}

For an ET evolution, an error occurring at a random instant $t$ should not affect the final output state except for an extra global phase, which can be represented by:
\begin{equation}
U(T,t)E_{j}U(t,0)|\psi_{\mathrm{L}}\rangle=e^{i\phi(t)}E_{j}U(T,0)|\psi_{\mathrm{L}}\rangle,\forall i,j,t.
\label{eq:ETcondition1}
\end{equation}
Here, $|\psi_{\mathrm{L}}\rangle$ is an arbitrary logical quantum state in the
code space, and $E_{j}$ is in the error set. Because Eq.~\ref{eq:ETcondition1} should be satisfied for arbitrary time $t$, it is equivalent to
\begin{equation}
U(t+\delta t,t)E_{j}|\psi_{\mathrm{L}}\rangle=e^{i\delta\phi(t)}E_{j}U(t+\delta t,t)|\psi_{\mathrm{L}}\rangle.\label{eq:ETcondition2}
\end{equation}
As a property of the logical gate, $U(t+\delta t,t)$ cannot introduce
leakage out of the code space, i.e. $U(t+\delta t,t)=\mathcal{P}_{\mathrm{C}}U(t+\delta t,t)\mathcal{P}_{\mathrm{C}}$ with $\mathcal{P}_{\mathrm{C}}$ being the projector onto the code space. Then the condition Eq.~\ref{eq:ETcondition2} can be transformed as
\begin{equation}
U(t+\delta t,t)E_{j}\mathcal{P}_{\mathrm{C}}|\psi_{\mathrm{L}}\rangle=e^{i\delta\phi(t)}E_{j}\mathcal{P}_{\mathrm{C}}U(t+\delta t,t)\mathcal{P}_{\mathrm{C}}|\psi_{\mathrm{L}}\rangle,
\label{eq:ETcondition3}
\end{equation}
by adding the projectors. By combining the requirements for QEC
\begin{equation}
\mathcal{P}_{\mathrm{C}}E_{j}^{\dag}E_{j}\mathcal{P}_{\mathrm{C}}=\alpha_{j}\mathcal{P}_{\mathrm{C}},\alpha_{j}\in\mathbb{R},\label{eq:ECcondition4}
\end{equation}
the ET condition becomes
\begin{equation}
\mathcal{P}_{\mathrm{C}}E_{j}^{\dag}U(t+\delta t,t)E_{j}\mathcal{P}_{\mathrm{C}}=e^{i\delta\phi(t)}\alpha_{j}\mathcal{P}_{\mathrm{C}}U(t+\delta t,t)\mathcal{P}_{\mathrm{C}}.\label{eq:ETcondition5}
\end{equation}
After introducing the projector $\mathcal{P}_{j}=\sqrt{\alpha_{j}}E_{j}\mathcal{P}_{\mathrm{C}}$ from the code space to the error space due to $E_{j}$, the above equation becomes
\begin{equation}
\mathcal{P}_{j}^{\dagger}U(t+\delta t,t)\mathcal{P}_{j}=e^{i\delta\phi(t)}\mathcal{P}_{\mathrm{C}}U(t+\delta t,t)\mathcal{P}_{\mathrm{C}}.
\label{eq:ETcondition6}
\end{equation}
Since $U\left(t+\delta t,t\right)=I-iH(t)\delta t+\mathcal{O}(\delta t^2)$,
the ET condition can also be represented by the system Hamiltonian
as
\begin{equation}
\mathcal{P}_{j}^{\dagger}H(t)\mathcal{P}_{j}=\mathcal{P}_{\mathrm{C}}H(t)\mathcal{P}_{\mathrm{C}}+c(t)\mathcal{P}_{\mathrm{C}},
\label{eq:ETcondition7}
\end{equation}
where $c(t)=-d\phi(t)/dt$. Here, the Hamiltonian
is presented in the code space.

In previous theoretical works~\citep{vy2013error,kapit2018error},
the condition of ET gates is derived as
\begin{equation}
[E_{j},H(t)]|\psi_{\mathrm{L}i}\rangle=0,\forall i,j,t.
\label{eq:wrongETcondition}
\end{equation}
Comparing with the ET condition (Eq.~\ref{eq:ETcondition7}) derived above, the commutation relation is too strict, and Eq.~\ref{eq:wrongETcondition} is just equivalent to a special case of Eq.~\ref{eq:ETcondition7} with $\phi(t)=0$. Therefore, the new ET condition provided in this work is more general with less restriction, and could relax the requirements for experimentally implementing the ET quantum gates.

%\bibliographystyle{Zou}
%\bibliography{ETgate}

%merlin.mbs apsrev4-1.bst 2010-07-25 4.21a (PWD, AO, DPC) hacked
%Control: key (0)
%Control: author (8) initials jnrlst
%Control: editor formatted (1) identically to author
%Control: production of article title (-1) disabled
%Control: page (0) single
%Control: year (1) truncated
%Control: production of eprint (0) enabled
%

\vspace{0.2in}